\documentclass[utf8]{frontiersFPHY_arXiv} 

\setcitestyle{square} 
\usepackage{url,hyperref,lineno,microtype,subcaption}
\usepackage[onehalfspacing]{setspace}

\usepackage{siunitx}

\DeclareSIUnit \dBm {dBm}

\usepackage{amsmath}
\DeclareMathOperator*{\argmin}{arg\,min}

\newcommand{\smallerrel}[1]{\mathrel{\mathpalette\smallerrelaux{#1}}}
\newcommand{\smallerrelaux}[2]{\raisebox{.1ex}{\scalebox{.75}{$#1#2$}}}

\def\firstAuthorLast{Pauwels {et~al.}} 
\def\Authors{Ja\"el Pauwels \,$^{1,2,*}$, Guy Verschaffelt\,$^{1}$, Serge Massar\,$^{2}$ and Guy Van der Sande\,$^{1}$}
\def\Address{$^{1}$Applied Physics Research Group, Vrije Universiteit Brussel, Brussels, Belgium \\
$^{2}$Laboratoire d'Information Quantique, Universit\'e libre de Bruxelles, Brussels, Belgium }
\def\corrAuthor{Ja\"el Pauwels}

\def\corrEmail{jael.pauwels@vub.be}

\usepackage{transparent}
\usepackage{draftwatermark}
\SetWatermarkText{preprint}
\SetWatermarkScale{1.2}

\begin{document}

\onecolumn
\firstpage{1}

\title[Distributed Nonlinearity in Optical Reservoir]{Distributed Kerr Nonlinearity in a Coherent All-Optical Fiber-Ring Reservoir Computer} 

\author[\firstAuthorLast ]{\Authors} 
\address{} 
\correspondance{} 

\extraAuth{}

\maketitle

\begin{abstract}

We investigate, both numerically and experimentally, the usefulness of a distributed nonlinearity in a passive coherent photonic reservoir computer. This computing system is based on a passive coherent optical fiber-ring cavity in which part of the nonlinearities are realized by the Kerr nonlinearity. Linear coherent reservoirs can solve difficult tasks but are aided by nonlinear components in their input and/or output layer. Here, we compare the impact of nonlinear transformations of information in the reservoir’s input layer, its bulk - the fiber-ring cavity - and its readout layer. For the injection of data into the reservoir, we compare a linear input mapping to the nonlinear transfer function of a Mach Zehnder modulator. For the reservoir bulk, we quantify the impact of the optical Kerr effect. For the readout layer we compare a linear output to a quadratic output implemented by a photodiode. We find that optical nonlinearities in the reservoir itself, such as the optical Kerr nonlinearity studied in the present work, enhance the task solving capability of the reservoir. This suggests that such nonlinearities will play a key role in future coherent all-optical reservoir computers.


\end{abstract}

\section{Introduction}

In this work, we discuss an efficient, i.e. high speed and low power, analogue photonic computing system based on the concept of reservoir computing (RC) \cite{maass2002,jaeger2004}. This framework allows to exploit the transient dynamics of a nonlinear dynamical system for performing useful computations. In this neuromorphic computing scheme, a network of interconnected computational nodes (called neurons) is excited with input data. The ensemble of neurons is called the reservoir, and the interneural connections are fixed and can be chosen at random. For the coupling of the input data to the reservoir an input mask is used: a set of input weights which determines how strongly each of the inputs couples to each of the neurons. The randomness in both the input mask and internal reservoir connections ensures diversity in the neural responses. 
The reservoir output is constructed through a linear combination of neural responses (possibly first processed by a readout function) with a set of readout weights. The strength of the reservoir computing scheme lies in the simplicity of its training method, where only the readout weights are tuned to force the reservoir output to match a desired target. In general, a reservoir exhibits internal feedback through loops in the neural interconnections. As a result any reservoir has memory, which means it can retain input data for a finite amount of time, and it can compute linear and nonlinear functions of the retained information. 

Within the field of reservoir computing two main approaches exist: in the network-based approach networks of neurons are implemented by connecting multiple discrete nodes \cite{verstraeten2007}, and in the delay-based approach networks of virtual neurons are created by subjecting a single node (often a nonlinear dynamical device) to delayed feedback \cite{appeltant2011}. In the latter, the neurons are called virtual because they correspond with the travelling signals found in consequent timeslots in the continuous delay-line system. On account of this time-multiplexing of neurons, the input weights are translated into a temporal input mask, which is mixed with the input data before it is injected into the reservoir. Besides ensuring diversity in the neural responses, this input mask also keeps the virtual neurons in a transient dynamic regime, which is a necessary condition for good reservoir computing performance. 

Multiple opto-electronic reservoirs have been implemented, both delay-based \cite{paquot2012,larger2012,duport2016,larger2017} and network-based \cite{bueno2018}. Several all-optical reservoirs have been realized, both network-based systems \cite{vandoorne2011,vandoorne2014,katumba2018,bueno2018,harkhoe2018} and delay-based systems \cite{duport2012,brunner2013,vinckier2015}. An overview of recent advances is given in Ref. \cite{vandersande2017}. We observe that in the field of optical reservoir computing, some implementations operated in an incoherent regime, while others operated in a coherent regime. Coherent reservoirs have the advantage that they can exploit the complex character of the optical field, exploit interferences, and can use the natural quadratic nonlinearity of photodiodes. As a drawback, coherent bulk optical reservoirs typically need to be stabilized, but this is not a problem for on chip implementations.
Here we investigate the potential advantage of having a coherent reservoir with nonlinearity inside the reservoir. We show that it can increase the performance of the reservoir on certain tasks and we expect that future coherent optical reservoir computers will make use of such nonlinearities. 

State of the art photonic implementations target simple reservoir architectures \cite{harkhoe2018}, which can easily be upscaled to increase the number of computational nodes or neurons, thereby enhancing the reservoir’s computational capacity. Even a linear photonic cavity can be a potent reservoir \cite{vinckier2015}, provided that some nonlinearity is present either in the mapping of input data to the reservoir, or in the readout of the reservoir’s response. Despite advances towards all-optical RC \cite{bienstman2018}, many state of the art photonic reservoir computers inherently contain some nonlinearity as they are usually set up to process and produce electronic signals. This means that even if the reservoir is all-optical, the reservoir computer in its entirety is of an opto-electronic nature. Commonly used components like a Mach-Zehnder modulators (MZM) and photodetectors (PD) provide means for transitioning back and forth between the electronic and optical domains, and they also – almost inevitably - introduce nonlinearities which boost the opto-electronic reservoir computer’s performance beyond the merits of the optical reservoir itself. When transitioning towards all-optical reservoir computers, such non-linearities can no longer be relied on, and thus the required nonlinear transformation of information must originate elsewhere. One option is then to use multiple strategically placed nonlinear components in the reservoir, but this can be a costly strategy when upscaling the reservoir \cite{vandoorne2011}. 

In this paper, we study a delay-based reservoir computer, based on a passive coherent optical fiber ring cavity following Ref. \cite{vinckier2015} and exploit the inherent nonlinear response of the waveguiding material to build a state-of-the-art photonic reservoir. This means that the nonlinearity of our photonic reservoir is not found in localized parts, but rather it is distributed over the reservoir’s entire extent. To correctly characterize the effects of such distributed nonlinearity, we also consider in this study all other nonlinearities that may surround the reservoir. In terms of the reservoir’s input mapping, we examined the system responses when receiving optical inputs (linear mapping), and when receiving electronic inputs coupled to the optical reservoir through a Mach-Zehnder modulator with a nonlinear mapping. For the reservoir’s readout layer, we examined both linear readouts (coherent detection) and nonlinear readouts through the quadratic nonlinearity of a photodiode measuring the power of the optical field. Taking these different options into account, we then constructed different scenarios in terms of the presence of nonlinearities in the input and/or output layer of these reservoir computers. In all these scenarios we numerically benchmarked the RC performance, thus quantifying the difference in performance between systems which do or do not have such distributed nonlinearity inside the reservoir. In the next sections, we show our numerical results, which show a broad range of optical input power levels at which these RCs benefit from the self-phase modulation experienced by the signals due to the nonlinear Kerr effect induced by the waveguide material. We also show the results of our experimental measurements that indicate how much this distributed nonlinearity boosts the reservoir's capacity to perform nonlinear computation. In the discussion section, we analyze the impact of these findings on the future of photonic reservoir computing.

\section{Materials and Methods}

\subsection{Setup} \label{section:setup}

\begin{figure}[h!]
\begin{center}
\includegraphics[width=8cm]{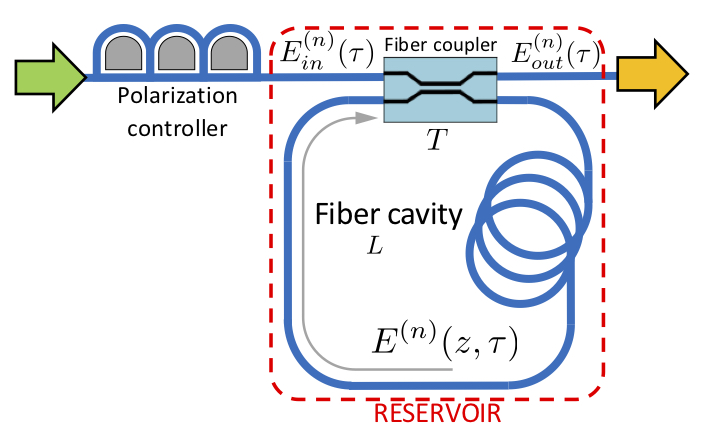}
\end{center}
\caption{Schematic of the fiber-ring cavity of length $L$ used to implement an optical reservoir. The green (orange) arrow indicates a connection with an input (output) layer. A polarization controller maps the input polarization onto a polarization eigenmode of the cavity. A coupler with power transmission coefficient $T$ couples the input field $E_{in}^{(n)}(\tau)$ to the cavity field $E^{(n)}(z,\tau)$ and couples to the output field $E_{out}^{(n)}(\tau)$, where $n$ is the roundtrip index, $\tau$ is time (with $0<\tau<t_R$) and $z$ is the longitudinal position in the ring cavity. }\label{fig:setup_core}
\end{figure}

Our reservoir computing simulations and experiments are based on the set of dynamical systems which are discussed in this section. The reservoir itself is implemented in the all-optical fiber-ring cavity shown in Fig. \ref{fig:setup_core}, using standard single-mode fiber. A polarization controller is used to ensure that the input field $E_{in}$ (originating from the green arrow) excites a polarization eigenmode of the fiber-ring cavity. A fiber coupler, characterized by its power transmission coefficient $T=50\%$, couples light in and out of the cavity. The fiber-ring is characterized by the roundrip length $L=\SI{10}{\meter}$ (or roundtrip time $t_R$), the propagation loss $\alpha$ (taken here \SI{0.18}{\deci\bel\per\kilo\meter}), the fiber nonlinear coefficient $\gamma$ (which is set to $0$ to simulate a linear reservoir, and set to $\gamma_{Kerr}=\SI{2.6}{\milli\radian\per\meter\per\watt}$ to simulate a nonlinear reservoir), and the cavity detuning $\delta_0$, i.e. the difference between the roundtrip phase and the nearest resonance (multiple of $2\pi$). This low-finesse cavity is operated off-resonance, with a maximal input power of \SI{50}{\milli\watt} (\SI{17}{\dBm}). A network of time-multiplexed virtual neurons is encoded in the cavity field envelope. The output field $E_{out}$ is sent to the readout layer (through the orange arrow) where the neural responses are demultiplexed.

\begin{figure}[h!]
\begin{center}
\includegraphics[width=8cm]{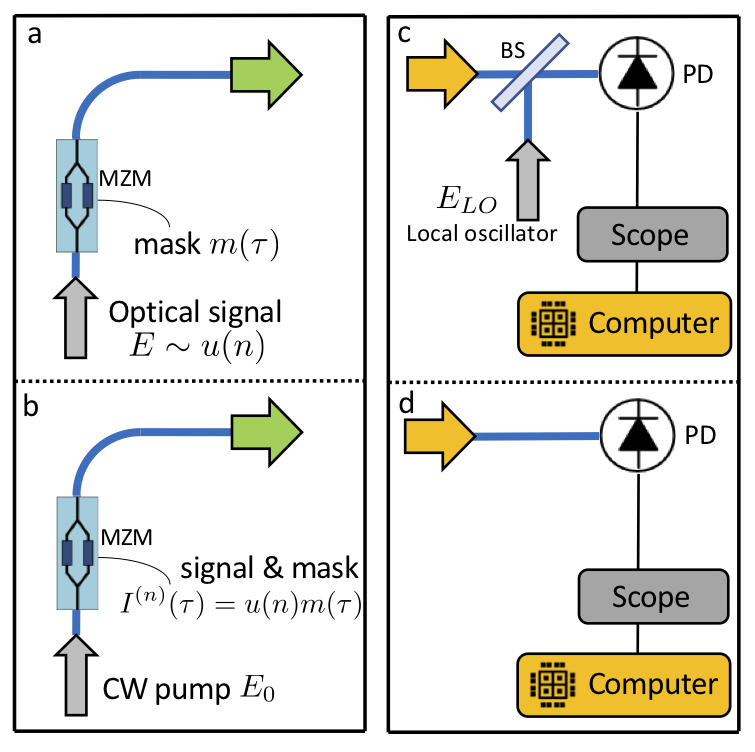}
\end{center}
\caption{Schematics of input and output layers connecting to the reservoir shown in Fig. \ref{fig:setup_core}. In the linear input scheme \textbf{(a)} the Mach-Zehnder modulator (MZM) superimposes the reservoir's input mask $m(\tau)$ on the optical signal $E\sim u(n)$ carrying the input data. In the (possibly) nonlinear input scheme \textbf{(b)} the input data is mixed with the input mask and then drives the MZM to modulate a CW optical pump. In the linear output scheme \textbf{(c)} a reference field $E_{LO}$ is used to implement coherent detection, allowing a quadrature of the complex optical field to be measured. Note that coherent detection requires two such readout arms with phase-shifted reference fields in order to measure the complex output field $E_{out}$. In the nolinear output scheme \textbf{(d)} only a photodetector (PD) is used, thus only allowing the optical output power $|E_{out}|^2$ to be recorded.} \label{fig:setup_inputs_outputs}
\end{figure}

The input field $E_{in}$ can originate from one of two different optoelectronic input schemes. Firstly we consider a scenario where the input signal $u(n)$ (with discrete time $n$) is amplitude-encoded in an optical signal $E\sim u(n)$, as shown in Fig. \ref{fig:setup_inputs_outputs}(a). The reservoir's input mask $m(\tau)$ is mixed with the input signal by periodic modulation of the optical input signal using an MZM. This scheme was implemented in Ref. \cite{duport2016}, but the nonlinearity of the MZM was avoided through pre-compensation of the electronic input signal. Note that the discrete time $n$ corresponds with the roundtrip index. And as delay-based reservoirs are typically set up to process 1 sample each roundtrip, $n$ also corresponds with the sample index. However, we have chosen to hold each input sample over multiple roundtrips, for reasons which are explained in the Results section (that is, $u(n)$ is constant over multiple values of $n$). Secondly we consider a scenario where we use the MZM to modulate a CW optical pump following Ref. \cite{duport2012}, as shown in Fig. \ref{fig:setup_inputs_outputs}(b). Here the input signal is first mixed with the input mask and then used to drive the MZM. It is known that the MZM's nonlinear transfer function can affect the RC system's performance \cite{vinckier2015}, but the implications for a coherent nonlinear reservoir have not yet been investigated. 

Similarly, the output field $E_{out}$ can be processed by two different  optoelectronic readout schemes. Firstly we consider a coherent detection scheme as shown in Fig. \ref{fig:setup_inputs_outputs}(c). Mixing the reservoir's output field with a reference field $E_{LO}$ allows to record the complex neural responses, time-multiplexed in the output field $E_{out}$. Secondly, we consider a readout scheme where a photodetector (PD) measures the optical power of the neural responses $|E_{out}|^2$, as shown in Fig. \ref{fig:setup_inputs_outputs}(d). 

With high optical power levels and small neuron spacing (meaning fast modulation of the input signal), dynamical and nonlinear effects other than the Kerr nonlinearity may appear, such as photon-phonon interactions causing Brillouin and Raman scattering, and bandwidth limitations caused by the driving and readout equipment. We want to focus in the present work on the effects of the Kerr nonlinearity. Combined with the memory limitations of the oscilloscope, we therefore limit our reservoir to $20$ neurons, with a maximal input power of \SI{100}{\milli\watt}.

The current setup is not actively stabilized. We have found that the cavity detuning $\delta_0$ does not vary more than a few \SI{}{\milli\radian} over the course of any single reservoir computing experiment, where a few thousand input samples are processed. A short header, added to the injected signal, allows us to recover the detuning $\delta_0$ post-experiment. We effectively measure the interference between a pulse which reflects off the cavity and a pulse which completes one roundtrip through the cavity. However, we find that the precise value of $\delta_0$ has no significant influence on the experimental reservoir computing results.

\subsection{Physical model}

Here we discuss the mean-field model used to describe the temporal evolution of the  electric field envelope $E^{(n)}(z,\tau)$ inside the cavity, where $n$ is the roundtrip index, $0<\tau<t_R$ is time (bound by the cavity roundtrip time $t_R$ and $0<z<L$ is the longitudinal coordinate of the fiber ring cavity with length $L$. The position $z=0$ corresponds to the position of the fiber coupler. The position $z=L$ corresponds to the same position, but after propagation through the entire fiber-ring. We will describe the evolution on a per-roundtrip basis (i.e. with varying roundtrip index $n$). With this notation $E^{(n)}(z,\tau)$ represents the cavity field envelope measured at position $z$ at time $\tau$ during the $n$-th roundtrip. For each roundtrip we model propagation through the nonlinear cavity to obtain $E^{(n})(z=L,\tau)$ from $E^{(n)}(z=0,\tau)$. We then express the cavity boundary conditions to obtain $E^{(n+1)}(0,\tau)$ from $E^{(n)}(L,\tau)$ and to obtain the field $E_{out}^{(n)}(\tau)$ at the output of the fiber-ring reservoir. For now we will omit $\tau$.

Firstly, to model propagation in the fiber-ring cavity we take into account propagation loss and the nonlinear Kerr-effect. Since the nonlinear propagation model is independent from the roundtrip index $n$, this subscript is omitted in the following description. The nonlinear propagation equation is given by 
\begin{equation} \label{eq:nonlinearpropagation}
\partial_z E = i\gamma |E|^2E-\alpha E.
\end{equation}
Here, $\alpha$ is the propagation loss and $\gamma$ is the nonlinear coefficient which is set to $\gamma=0$ to simulate a linear reservoir, and set to $\gamma=\gamma_{Kerr}$ to include the nonlinear Kerr effect caused by the fiber waveguide. We do not include dispersion effects at the current operating point of the system, since the neuron separation is much larger than the diffusion length, hence also $\tau$ can be omitted in the nonlinear propagation model. The evolution of the power $|E(z)|^2$ is readily obtained by solving the corresponding propagation equation 
\begin{equation} \label{eq:powerequation}
\partial_z |E|^2 = E^* \partial_z E  + E \partial_z E^* = -2\alpha |E|^2,
\end{equation}
\begin{equation} \label{eq:powersolution}
|E(z)|^2 = |E(0)|^2 e^{-2\alpha z}.
\end{equation}
With $\phi_{_z}$ the nonlinear phase acquired during propagation over a distance $z$, we know that the solution of $E(z)$ will be of the form
\begin{equation} \label{eq:fieldsolutionformal}
E(z) = E(0) e^{i\phi_{_z}-\alpha z}.
\end{equation}
Since this nonlinear phase depends on the power evolution given by Eq. \eqref{eq:powersolution}, an expression for $\phi_{_z}$ is found to be
\begin{equation} \label{eq:nonlinearphase}
\phi_{_z} = \gamma \int_0^z |E(v)|^2\delta v = \gamma |E(0)|^2 \int_0^z e^{-2\alpha v}\delta v = \gamma |E(0)|^2 \frac{1-e^{-2\alpha z}}{2 \alpha}.
\end{equation}
At this point, we can introduce the effective propagation distance $z_{eff}$ as
\begin{equation} \label{eq:effectivedistance}
z_{eff} = \frac{1-e^{-2\alpha z}}{2 \alpha}.
\end{equation}
In general (since $\alpha\geq0$) we have $z_{eff}\leq z$.
Substituting these result in Eq. \eqref{eq:fieldsolutionformal} yields the complete solution for propagation of the cavity field envelope 
\begin{equation} \label{eq:fieldsolutionfull}
E(z) = E(0) \exp\left(i\gamma |E(0)|^2 z_{eff}-\alpha z\right).
\end{equation}

Finally, we reinstitute the roundtrip index $n$ and the time parameter $\tau$ which allows us to combine this nonlinear propagation model with the cavity boundary conditions.
\begin{align} \label{eq:fullmodel}
\left \{ \begin{array}{rl} E^{(n)}(L,\tau) &= E^{(n)}(0,\tau) \exp\left(i\gamma |E^{(n)}(0,\tau)|^2 L_{eff}-\alpha L\right) \\ E^{(n+1)}(0,\tau) &= \sqrt{T}E_{in}^{(n+1)}(\tau) + \sqrt{1-T}e^{i\delta_{_0}}E^{(n)}(L,\tau) \\ E_{out}^{(n+1)}(\tau) &= \sqrt{1-T}E_{in}^{(n+1)}(\tau) + \sqrt{T}e^{i\delta_{_0}}E^{(n)}(L,\tau) \end{array} \right.
\end{align}
In these equations, $T$ represents the power transmission coefficient of the cavity coupler, and $\delta_0$ represents the cavity detuning (i.e. difference between the roundtrip phase and the closest cavity resonance). Further, the input field $E_{in}=E_{in}^{(n)}(\tau)$ changes with the roundtrip index $n$ as new data samples can be injected into the system, and is modulated in time using the input mask to create a network of virtual neurons. The output field $E_{out}=E_{out}^{(n)}(\tau)$ containing the neural responses is sent to a measurement stage.

\subsection{Reservoir computing}

The framework of reservoir computing allows to exploit the transient nonlinear dynamics of a dynamical system to perform useful computation \cite{maass2002,jaeger2004}.
For the purpose of reservoir computing, virtual neurons (dynamical variables, computational nodes) are time-multiplexed in $\tau$-space of the physical system described by Eq. \eqref{eq:fullmodel}, following the delay-based reservoir computing scheme originally outlined in Ref. \cite{appeltant2011}. As such, the input field $E_{in}^{(n)}(\tau)$ varies with $n$ as new input samples arrive, and varies with $\tau$ to implement the input mask, which excites the neurons into a transient dynamic regime. Subsequently, the neural responses are encoded in the output field $E_{out}^{(n)}(\tau)$ and need to be demultiplexed from $\tau$-space.
As in Refs. \cite{paquot2012,vinckier2015} the length $t_M$ of the input mask $m(\tau)$ is deliberately mismatched from the cavity roundtrip time $t_R$. Instead, we set $t_M = t_R N / (N+1)$ which provides interconnectivity between the $N$ virtual neurons in a ring topology. The input mask $m(\tau)$ is a piecewise constant function, with intervals of duration $\theta = t_M/N$. The signal $I^{(n)}(\tau)$ injected into the RC is constructed by multiplying the input series $u(n)$ with the input mask, $I^{(n)}(\tau) = u(n)m(\tau)$. When the input is coupled linearly to the reservoir then $E_{in}^{(n)}(\tau) \sim I^{(n)}(\tau)$. This would be the case when $u(n)$ is an optical signal periodically modulated with the input mask signal $m(\tau)$. When a MZM modulator with transfer function $f$ is used to convert the electronic signal $I^{(n)}(\tau)$ to the optical domain then $E_{in}^{(n)}(\tau) \sim f(I^{(n)}(\tau))$, where $f$ can be nonlinear. 

\begin{figure}[h!]
\begin{center}
\includegraphics[width=\linewidth]{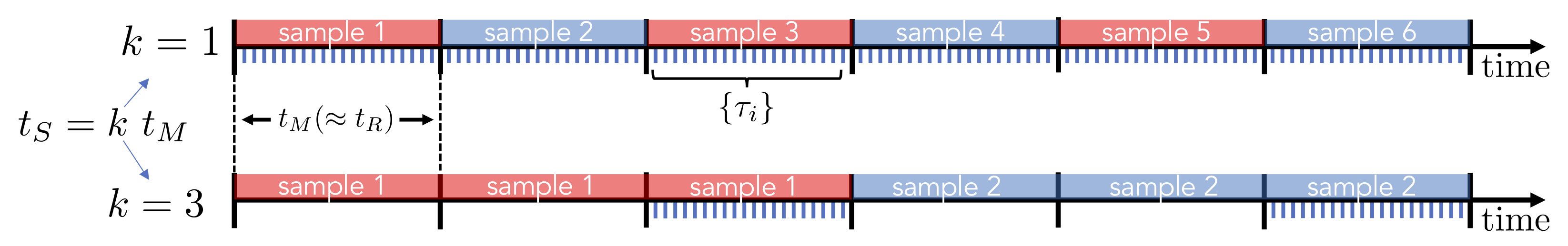}
\end{center}
\caption{Schematic of input and output timing, with $t_S$ the sample duration, $t_M$ the input mask duration and $t_R$ the roundtrip time. Input samples are injected during (integer) $k$ roundtrips (bars in alternating colors) and the neural responses are recorded at times $\{\tau_i\}$ (blue tick marks) during the last of those $k$ roundtrips.} \label{fig:timing}
\end{figure}

Note that in Ref. \cite{vinckier2015} the sample duration $t_S$ is matched to the length of the input mask $t_M$, allowing the reservoir to process 1 input sample approximately every roundtrip, as $t_S=t_M\smallerrel{\lesssim} t_R$. However, for reasons explained in the Results section, we will study different sample durations by holding input samples over multiple durations of the input mask, $t_S=k\ t_M$ with integer $k$ as illustrated in Fig. \ref{fig:timing}. This inevitably slows the reservoir down, as it only processes 1 input sample approximately every $k$ roundtrips. But it also provides practically straightforward means to accumulate more nonlinear processing of the data inside the reservior, which can then be measured and quantified.

Since the virtual neurons are time-multiplexed in this delay-based reservoir computer, they need to be de-multiplexed from $E_{out}^{(n)}(\tau)$ in the readout layer by sampling this output field at a set of times $\{\tau_i\}$ (with $i$ the neuron index and $1<i<N$ when $N$ neurons are used) as shown in Fig \ref{fig:timing}. The dynamical neural responses $x_i(n) = E_{out}^{(n)}(\tau_i)$ are recorded and used to train the reservoir to perform a specific task. That is, we optimize a set of readout weights $w_i$ which are used to combine the neural readouts into a single scalar reservoir output $y(n)$. In general the reservoir output is constructed as
\begin{equation}\label{eq:RCoutput}
y(n) = \sum_{i=1}^N w_i g(x_i(n))
\end{equation}
where the neural responses $x_i(n)$ are first parsed by an output function $g(x)$ taking into account the operation of the readout layer and readout noise $\nu$. In all simulations the fixed level of readout noise is matched to the experimental conditions. When the complex-valued reservoir states are directly recorded, then $g(x) = x+\nu$ and the readout weights $w_i$ are complex too, such that $y$ is real. If however, a PD measures the power of the neural responses, then $g(x)=|x|^2+\nu$ which is real-valued, and the readout weights will be real-valued too.
Tasks are defined by the real-valued target output $\hat{y}$.
Optimization of the readout weights occurs over a training set of $T_{train}$ input and target samples, and is achieved through least squares regression. This procedure minimizes the mean squared error between the reservoir output $y$ and target output $\hat{y}$, averaged over all samples.
\begin{equation} \label{eq:regression}
\{w_i\} = \argmin_{\{w_i\}} \langle \left( \hat{y} -  \sum_{i=1}^N w_i g(x_i) \right)^2 \rangle_{_{T_{train}}}.
\end{equation}
These optimized readout weights are then validated on a test set of $T_{test}$ new input and target samples. A common figure of merit to quantify the reservoir's performance is the normalized mean square error (NMSE) defined as 
\begin{equation}
NMSE(y,\hat{y}) = \frac{\langle \left( y-\hat{y} \right)^2 \rangle_{_{T_{test}}}}{\langle \hat{y} ^2 \rangle_{_{T_{test}}}}.
\end{equation}

\subsection{Balanced Mach-Zehnder modulator operation} \label{section:MZM}
Here we briefly investigate the relevant nonlinearities which occur when mapping an electronic signal to an optical signal using an MZM. The operation of our balanced MZM can be described as
\begin{equation} \label{eq:MZM}
\frac{E_{in}}{E_0} = \cos\left(\frac{V}{V_{\pi}}\frac{\pi}{2}\right)
\end{equation}
where $E_0$ represents the incident CW pump field, $E_{in}$ is the transmitted field which will be the input field to the optical reservoir, $V_{\pi}$ determines at which voltage the zero intensity point occurs (point of no transmission), and $V$ is the voltage of the applied electronical signal consisting of a bias contribution $V_{b}$ and a zero-mean signal $V_s$, i.e. $V = V_{b}+V_{s}$. For our numerical investigation, we will set the amplitude of the signal voltage to $|V_s| = V_{\pi}/2$. First, we investigate the zero intensity bias point, $V_b=V_{\pi}$. In this case, we can approximate Eq. \eqref{eq:MZM} with the following Taylor expansion
\begin{align}
\frac{E_{in}}{E_0} &= f(V_s) + \mathcal{O}\left(V_s^5 \right) \\
f(V_s) &= -\frac{\pi}{2V_{\pi}}V_s +\frac{1}{6} \left( \frac{\pi}{2V_{\pi}}\right)^3 V_s^3 \label{eq:taylorexpansion1}
\end{align}
With $\left(E_{in}/E_0\right)_{max}$ representing the maximal value of $\frac{E_{in}}{E_0}$ with the given bias voltage $V_b$ and signal amplitude $|V_s|$, the relative error $r.e.$ of the Taylor expansion \eqref{eq:taylorexpansion1}
\begin{equation} \label{eq:relativeerror}
r.e. = \frac{|\frac{E_{in}}{E_0}-f(V_s) |}{\left(\frac{E_{in}}{E_0}\right)_{max}}
\end{equation}
is smaller than $1\%$. When the cubic term ($\sim V_s^3$) of the approximation $f(V_s)$ is omitted, this error increases to $11\%$. This means that at this operating point of the MZM, there is a significant nonlinearity which scales with the input signal cubed.

Next, we investigate the linear intensity operating point, $V_b = V_{\pi}/2$. Although the MZM's transfer function at this operating point is the most linear in terms of the transmitted optical power, it is highly nonlinear in terms of the transmitted optical field. In this case, we replace Eq. \eqref{eq:taylorexpansion1} with
\begin{equation}
f(V_s) = \frac{1}{\sqrt{2}} \left(1-\frac{\pi}{2V_{\pi}}V_s + \frac{1}{2} \left( \frac{\pi}{2V_{\pi}}\right)^2 V_s^2 + \frac{1}{6} \left( \frac{\pi}{2V_{\pi}}\right)^3 V_s^3 + \frac{1}{24} \left( \frac{\pi}{2V_{\pi}}\right)^4 V_s^4 \right),
\end{equation} 
as we need all polynomial terms up to order 4 to keep the relative error defined by Eq. \eqref{eq:relativeerror} below $1\%$.
In this case, omitting terms of orders above 1 in the approximation $f(V_s)$ increases the relative error of the Taylor expansion to $26\%$. This means that at this operating point of the MZM there are multiple polynomial nonlinearities and that the total nonlinear signal distortion is stronger compared with the zero intensity bias point.

Furthermore, during our experiments we have decided to operate the MZM in a linear regime. This allows for the nonlinear effects inside the reservoir to be more readily measured. To this end, we tuned the MZM close to the zero intensity operating point, $V_b=V_{\pi}-\delta_V$ with $\delta_V \ll V_{\pi}$ and reduced the signal amplitude $|V_s|$. The small deviation $\delta_V$ is used to generate a bias in the optical field injected into the reservoir.

\subsection{Memory capacities} \label{section:memorycapacities}

To benchmark the performance of an RC, one can train it to perform one or several benchmark tasks. Alternatively, there exists a framework to quantify the system's total information processing capacity. This capacity is typically split into two main parts: the capacity of the system to retain past input samples is captured by the linear memory capacity \cite{jaeger2002}, and the capacity of the system to perform nonlinear computation is captured by the nonlinear memory capacity\cite{dambre2012}. It is known that the total memory capacity has an upper bound given by the number of dynamical variables in the system, which in our system is the number of neurons in the reservoir. It is also known that readout noise reduces this total memory capacity, and that there is a trade-off between linear and nonlinear memory capacity, depending on the operating regime of the dynamical system.
In order to measure these capacities for our reservoir computer a series of independent and identically distributed input samples $u(n)$ drawn uniformly from the interval $[-1,1]$ is injected into the reservoir, with discrete time $n$. The RC is subsequently trained to reconstruct a series of linear and nonlinear polynomial functions depending on past inputs $u(n-i)$, looking back $i$ steps in the past. Following Ref. \cite{dambre2012} these functions are chosen to be Legendre polynomials $P_d(u)$ (of degree $d$), because they are orthogonal over the distribution of the input samples. As an example, we can train the reservoir to reproduce the target signal $\hat{y}(n)$, given by
\begin{equation} \label{eq:memorytask}
\hat{y}(n) = P_2(u(n-1))P_1(u(n-3)).
\end{equation}
The ability of the RC to reconstruct each of these functions is evaluated by comparing the reservoir's trained output $y$ with the target $\hat{y}$ for previously unseen input samples. This yields a memory capacity $C$ which lies between $0$ and $1$ \cite{dambre2012},
\begin{equation} \label{eq:memorycapacity}
C = 1-\frac{\langle \left(\hat{y}-y\right)^2 \rangle}{\langle \hat{y}^2 \rangle},
\end{equation}
where $\langle . \rangle$ denotes the average over all samples used for the evaluation of $C$. Due to the orthogonality of the polynomial functions over the distribution of the input samples, the capacities corresponding to different functions yield independent information and can thus be summed to quantify the total memory capacity, i.e. the total information processing capacity of the RC. The memory functions are typically grouped by their total degree, which is the sum of degrees over all constituent polynomial functions, e.g. Eq. \eqref{eq:memorytask} has total degree 3. Summing all memory capacities corresponding with functions of identical total degree yields the total memory capacity per degree. This allows to quantify the contributions of individual degrees to the total memory capacity of the RC, which is the sum over all degrees. As the memory capacities will become small for large degrees, the total memory capacity is still bound. 

Since the reservoirs are trained and their performance is evaluated on finite data sets, we run the risk of overestimating the memory capacities $C$, whose estimator Eq. \eqref{eq:memorycapacity} is plagued by a positive bias \cite{dambre2012}. Therefore, a cutoff capacity $C_{co}$ is used ($C_{co}\approx 0.1$ for 1000 test samples) and capacities below this cutoff are neglected (i.e. they are assumed to be 0).

Note that the trade-off between linear and nonlinear memory capacity is typically evaluated by comparing the total memory capacity of degree 1 (linear) with the total memory capacity of all higher degrees (nonlinear). However, special attention is due when a PD is present in the readout layer of our RC. If a reservoir can (only) linearly retain past inputs $u(n-i)$ ($i$ steps in the past) then any neural response $x(n)$ consists of a linear combination (with a bias term $b$ and fading coefficients $a_i$) of those past inputs
\begin{equation} \label{eq:linearreservoirstate}
x(n) = b + \sum_i a_iu(n-i)
\end{equation}
and subsequently the optical power $P_x$ measured by the PD is given by
\begin{equation} \label{eq:PD}
P_x(n) = x(n)\bar{x}(n)  =|b|^2 + \sum_{i}2\text{Re}(b\bar{a_i})u(n-i)\  + \sum_{i,j}2\text{Re}(a_i\bar{a_j})u(n-i)u(n-j)
\end{equation}
which consists of polynomial functions of past inputs of degree 1 and 2.
Thus, in this case the total linear memory capacity of the RC is represented by the total memory capacity of degrees 1 and 2 combined. In case the bias term $b$ is lacking, only memory capacities of degree 2 will be present. On the other hand, if a PD is used in the output and memory capacities of degree higher than 2 are present, then this indicates that the reservoir itself is not linear, i.e. cannot be represented by a function of the form Eq. \eqref{eq:linearreservoirstate}.

\section{Results}

\subsection{Numerical RC performance: Sante Fe time series prediction}
For the injection of input samples to the optical reservoir, we consider two strategies as discussed in Section \ref{section:setup} and in Figs. \ref{fig:setup_inputs_outputs}(a) and (b), referred to here as the linear and nonlinear input regimes respectively. The exact shape of the nonlineariy in the nonlinear regime depends, among other things, on the operating point (or bias voltage) of the MZM, as discussed in Section \ref{section:MZM}. We will demonstrate this by showing results around both the linear intensity operating point and the zero intensity operating point of the MZM. For the readout of the reservoir response, we also consider two cases as discussed in Section \ref{section:setup} and in Figs. \ref{fig:setup_inputs_outputs}(c) and (d), referred to here as the linear and nonlinear output regimes respectively.

\begin{figure}[h!]
\begin{center}
\includegraphics[width=13.5cm]{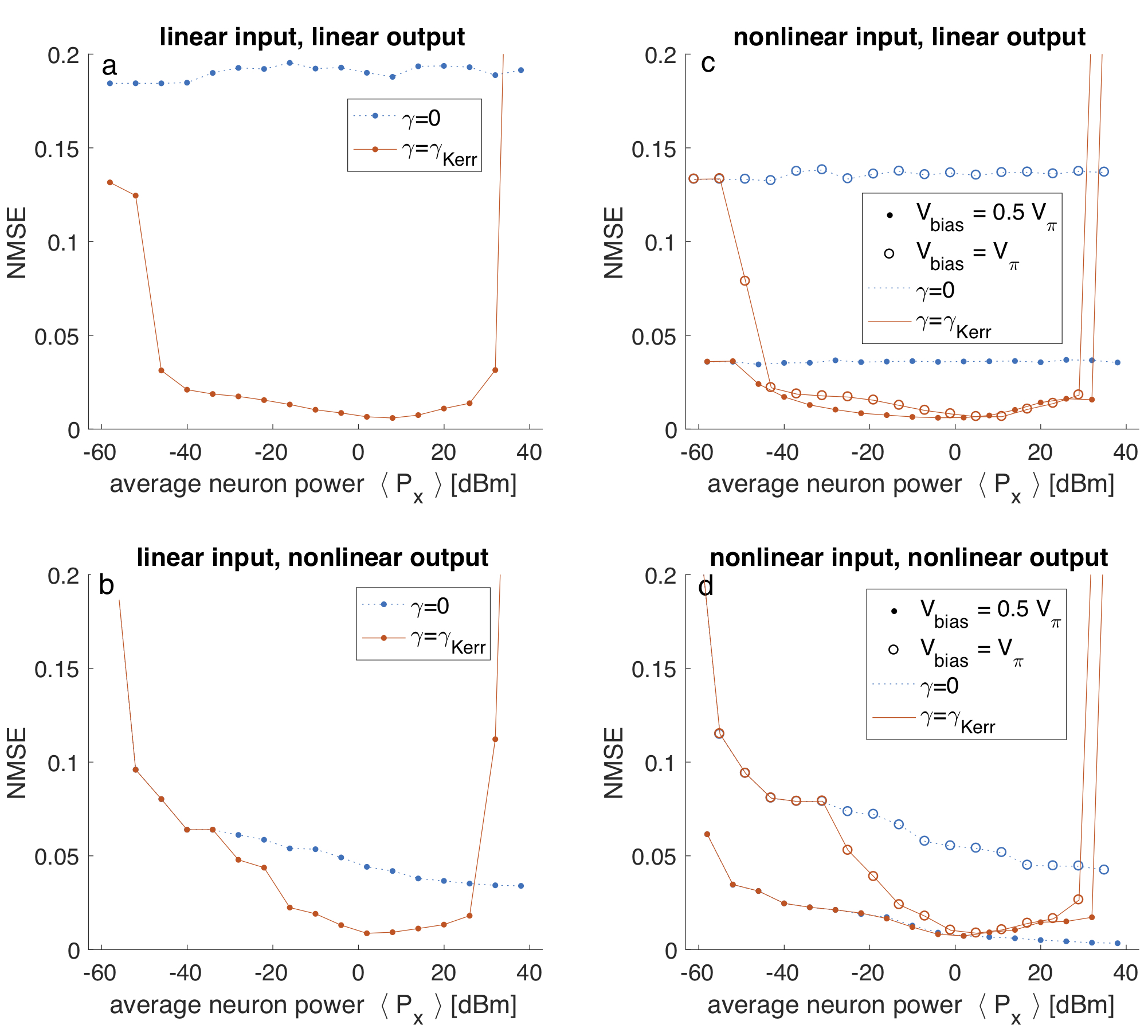}
\end{center}
\caption{Numerical results of fiber-ring reservoir computer on Santa Fe time series prediction tasks. In all panels the prediction error (NMSE) is plotted versus the average neuron power $\langle P_x \rangle$. Panels \textbf{(a)} and \textbf{(c)} correspond with a linear input layer, where panels \textbf{(b)} and \textbf{(d)} correspond with a nonlinear input layer using the MZM's nonlinear transfer function. The nonlinear input regime shows results for 2 different operating points of the MZM with different strengths of nonlinear transformation. Panels \textbf{(a)} and \textbf{(b)} correspond with a linear output layer, where panels \textbf{(c)} and \textbf{(d)} correspond with a nonlinear output layer using the PD.}\label{fig:numericalresults}
\end{figure}

We have thus identified 4 different scenarios based on the absence or presence of nonlinearities in the input and output layer of the reservoir computer. As we will show, we have for each of these cases numerically investigated the effect of the distributed nonlinear Kerr effect, present in the fiber waveguide, on RC performance. For this evaluation, we have used $100$ neurons to solve the Santa Fe time series prediction task \cite{weigend1993} and each input sample is injected during 6 roundtrips ($t_S=kt_M$ with $k=6$) for reasons which will become clear in Section \ref{section:experimentalresults}. Here, a pre-existing signal generated by a laser operating in a chaotic regime is injected into the reservoir. The target at each point in time is for the reservoir computer to predict the next sample. Performance is evaluated using the NMSE, where lower is better. Fig. \ref{fig:numericalresults} has 4 panels corresponding to these 4 scenario's. Each panel shows the NMSE as function of the average optical power per neuron inside the cavity. Dashed blue lines correspond with simulation results of linear reservoirs (i.e. with the nonlinear coefficient $\gamma$ set to $0$), and full red lines correspond with simulation results of reservoirs with Kerr-nonlinear waveguides (i.e. $\gamma$ set to $\gamma_{Kerr}$).

In Fig. \ref{fig:numericalresults}(a) both the input and output layers of the reservoir are strictly linear (i.e. optical input and coherent detection). It is clear that the linear reservoir ($\gamma=0$) scores poorly, with the NMSE approaching $20\%$. For a wide range of optical power levels, the presence of the Kerr nonlinear effect ($\gamma=\gamma_{Kerr}$) induced by the fiber waveguide boosts the RC performance, with an optimal NMSE just below $1\%$. This can be readily understood as it is well known that for this task, some nonlinearity is required in order to obtain good RC performance. Note that the average neuron power $\langle P_x \rangle$ can be used to estimate the average nonlinear phase $\phi_{Kerr}$ the signals will acquire during the sample duration $t_S$, as $\phi_{Kerr}=\gamma_{Kerr}\langle P_x \rangle L t_S/t_M$.
We observe that without the presence of phase noise in the cavity, the boost to the RC performance due to the Kerr effect starts at very small values of the estimated nonlinear phase, and breaks down when $\phi_{Kerr} \gtrsim 1$. Switching to Fig. \ref{fig:numericalresults}(b) we have now introduced the square nonlinearity by using a PD in the readout layer. Focusing on the results obtained with a linear reservoir, we see that the PD's nonlinearity alone decreases the NMSE down from $20\%$ to approximately $5\%$ ($\gamma=0$). Although the PD's nonlinearity clearly boosts the RC performance on this task, its effect is rather restricted. The PD only generates squared terms, and linear terms if a bias is present, see Section \ref{section:memorycapacities}, depending on the MZM's operating point. Furthermore this nonlinearity does not affect the neural responses nor the operation of the reservoir itself, as it only applies to the readout layer. It can thus be understood that the introduction of the Kerr nonlinearity inside the reservoir warrants an additional significant drop in NMSE, to below $1\%$ ($\gamma=\gamma_{Kerr}$).
In Fig. \ref{fig:numericalresults}(c), the output layer is linear again, but now we have introduced the MZM in the input layer. 
The closed markers correspond with simulations where the MZM operates around the zero intensity operating point or the point of minimal transmission ($V_{bias}=V_{\pi}$). In terms of the optical field modulation, this is the most linear regime. It is thus no surprise that the performance of both linear and nonlinear reservoirs mimics that Fig. \ref{fig:numericalresults}(a) where no nonlinearity was present in the input layer. The only difference is that the error of the linear reservoir drops from $20\%$ to about $13\%$  ( $\gamma=0$, $V_{bias}=V_{\pi}$) because of the small residual nonlinearity at this operating point of the MZM. 
The round markers correspond with simulations where the MZM operates around the linear intensity operating point  ($V_{bias}=V_{\pi}/2$). In terms of the optical field modulation, the nonlinearity in the mapping of input samples to the optical field injected into the reservoir is more nonlinear at this operating point. This is why even the linear reservoir  manages to achieve errors below $4\%$ ($\gamma=0$, $V_{bias}=V_{\pi}/2$). Again we see that the introduction of the nonlinear Kerr effect allows the NMSE to drop even further, to below $1\%$ ($\gamma=\gamma_{Kerr}$). In fact, this scenario is similar to the scenario with linear input mapping and nonlinear output mapping, Fig. \ref{fig:numericalresults}(b). 
Finally, in Fig. \ref{fig:numericalresults}(d), nonlinearities are present in both the input mapping and readout layer. With the MZM operating around the zero intensity operating point, there is only a weak nonlinearity in the input mapping and thus, as expected, both linear and nonlinear reservoirs show trends which are very similar to the scenario where the input mapping is linear, Fig. \ref{fig:numericalresults}(c). With the MZM operating around the linear intensity operating point ($V_{bias}=V_{\pi}/2$) however, we observe a scenario in which the RC does not seem to benefit from the presence of the Kerr nonlinear effect. It seems that with significant nonlinearities present in both input and output layers of the RC the distributed nonlinear effect inside the reservoir cannot further descrease the NMSE below values attained by the linear reservoir, which is below $1\%$ ($V_{bias}=V_{\pi}/2$). In all other cases, Figs. \ref{fig:numericalresults}(a,b,c), we find that the distributed nonlinearity inside the reservoir significanlty boosts RC performance, and we find that its presence is critical when no other nonlinearities are available.

\subsection{Experimental verification: linear and nonlinear memory capacity} \label{section:experimentalresults}

In this Section we compare experimental results with detailed numerical simulations. For the experimental verification of our work, we are currently limited to operate with 20 neurons, as explained in Section \ref{section:setup}. Therefore, we have chosen not to perform the reservoir computing experiment on the Santa Fe task. With this few neurons, tasks like the Santa Fe task become hard for the reservoir. Instead we turn to a more academic task which allows us to quantify the reservoir's memory and nonlinear computational capacity in a more complete and task-independent way. We experimentally measure the linear and nonlinear memory capacities considered in Section \ref{section:memorycapacities}. Even with this few neurons the evaluation of the memory capacities can yield meaningful results while taking up comparatively little processing time.

\begin{figure}[h!]
\begin{center}
\includegraphics[width=15cm]{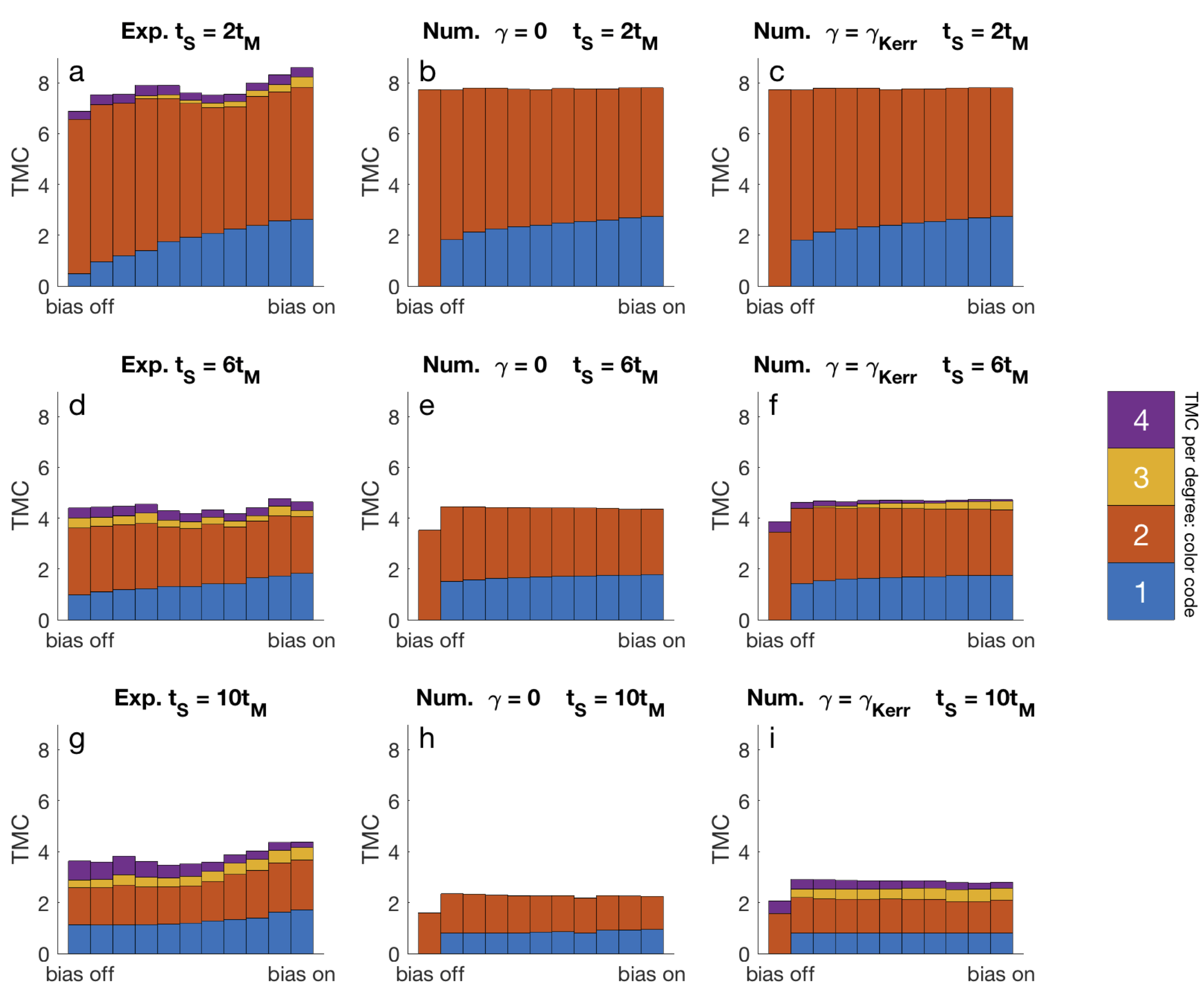}
\end{center}
\caption{Comparison between experimental results \textbf{(a,d,g)} and numerical models with linear ($\gamma=0$) \textbf{(b,e,h)} and nonlinear ($\gamma=\gamma_{Kerr}$) reservoirs \textbf{(c,f,i)}. The stacked vertical bars are color-coded to respresent the total memory capacities (TMC) of degree 1 (blue), 2 (red), 3 (orange), and 4 (purple). As such, the total height represents the total overall memory capacity. A control variable to the MZM $\delta_V$, is varied to include a small bias component to the injected optical field, where \textit{bias off} corresponds with $\delta_V=0$ and \textit{bias on} corresponds with a small nonzero value $0<\delta_V\ll V_{\pi}$. The sample duration $t_S$ is varied from 2 times  \textbf{(a,b,c)}, to 6 times \textbf{(d,e,f)} and finally to 10 times \textbf{(g,h,i)} the input mask period $t_M$ ($\approx$ cavity roundtrip time $t_R$).}\label{fig:experimentalverification}
\end{figure}

For these experiments, the input layer to our fiber-ring reservoir contains a balanced MZM tuned to operate in a linear regime as outlined in Section \ref{section:MZM}. The output layer employs a PD to measure the neural responses. That is, we use the setups of Figs. \ref{fig:setup_inputs_outputs}(b) and (d) but with the MZM operated as in Eq. \eqref{eq:taylorexpansion1}. Following Ref. \cite{dambre2012}, we have driven the reservoir with a series of independent and identically distributed random samples and trained the RC to reproduce different linear and nonlinear polynomial functions of past input samples. The capacity of the reservoir to reconstruct these functions was then evaluated and results were grouped according to the function's polynomial degree. To retain oversight on the results, we will only show the total capacity per degree, by summing all capacities corresponding with functions of the same total polynomial degree. In Fig. \ref{fig:experimentalverification} we show the total memory capacity per degree, encoded in the height of vertically stacked and color-coded bars. The stacking allows to visualize the contributions of individual degrees to the total overall memory capacity (summed over all degrees). Capacities of degree higher than 4 are not considered, as they were found not to contribute significantly to the total memory capacity of the system. 
For results labeled \textit{bias off} the MZM operates at the zero-intensity point ($V_{bias}=V_{\pi}$), and moving towards the \textit{bias on} label, we tuned the MZM's bias voltage ($V_{bias}=V_{\pi}-\delta_V$, with $\delta_V \ll  V_{\pi}$). This introduces a small bias component to the optical field injected into the reservoir, without compromising the linear operation of the MZM. 
The experiment was also repeated for different values of the sample duration $t_S$ with respect to the input mask periodicity $t_M$ (approximately equal to the cavity roundtrip $t_R$). We expect the sample duration to play a very important role, since it determines how much time a piece of information spends inside the cavity, and thus how much nonlinear phase can be acquired. The ratio $t_S/t_M$ is gradually increased from $t_S=2t_M$ in (first row) Figs. \ref{fig:experimentalverification}(a), (b) and (c), to  $t_S=6t_M$ in (middle row) Figs. \ref{fig:experimentalverification}(d), (e) and (f), and finally to  $t_S=10t_M$ in (bottow row) Figs \ref{fig:experimentalverification}(g), (h) and (i).
The experimental results in (left column) Figs. \ref{fig:experimentalverification}(a),(d) and (g) are compared with numerical results on a linear reservoir ($\gamma=0$) in (middle column) Figs. \ref{fig:experimentalverification}(b), (e) and (h), and a nonlinear reservoir ($\gamma=\gamma_{Kerr}$) in (right column) Figs. \ref{fig:experimentalverification}(c), (f) and (i). 

Firstly, in Fig. \ref{fig:experimentalverification}(a) we observe that without bias to the optical input field ($V_{bias}=V_{\pi}$) the total memory capacity originates almost completely from the polynomial functions of degree 2 which means (given the presence of the PD in the readout layer) that the optical system is almost completely linear.
Then, as an optical field bias is introduced we find that the total linear memory capacity of the system is now shared between degrees 1 and 2. As expected on account of quadratic nonlinearity due to the PD, Eq. \eqref{eq:PD}, the contribution of (odd) degree 1 grows with the increasing bias. Beyond these capacities of degree 1 and 2, we also observe a small contribution of capacities of degrees 3 and 4. We ascribe these contributions to the imperfect tuning of the MZM and thus a small residual nonlinearity in the input mapping. Note that the simulations take into account the quasi-linear input mapping of the MZM, but seemingly underestimate the residual nonlinearities to be insignificant. The imperfection of the MZM tuning also leads to a small residual bias component to the optical injected field, resulting in a small non-zero capacity of degree 1. Numerical simulations of linear ($\gamma=0$) and nonlinear ($\gamma=\gamma_{Kerr}$) reservoirs in Figs. \ref{fig:experimentalverification}(b) and (c) respectively, show the same growth in the memory capacity of degree 1 at the expense of the memory capacity of degree 2 when the bias is changed. Note that both simulations seem to overestimate the minimal bias required to obtain a significant memory capacity of degree 1. At this sample duration ($t_S=2t_M$) neither simulations indicate any significant contributions of capacities with degrees beyond 2. 

When increasing the sample duration ($t_S=6t_M$ and $t_S=10t_M$), the experimenal results in Figs. \ref{fig:experimentalverification}(d) and (g) show a steady increase in the contributions of capacities with degrees 3 and 4. This increase is attributed to the nonlinear Kerr effect, due to the larger accumulation of nonlinear phase during the time each sample is presented to the reservoir. At the same time we see a decrease in the capacities of degrees 1 and 2. As explained before, due to the PD these capacities capture the reservoir's capacity to linearly retain past samples. This trade-off between linear memory capacity (here degrees 1 and 2) and nonlinear computational capacity (here degrees 3 and 4) is well documented \cite{dambre2012}. Because we use the sample duration ($t_S=kt_M\approx kt_R$) to control the cumulative nonlinear effect inside the reservoir, we inevitably increase the mismatch between the inherent timescale of the input data (i.e. the sample duration $t_S$) and the inherent timescale of the reservoir (i.e. the cavity roundtrip $t_R$). and alter the reservoir’s internal topology. When each sample is presented longer, past samples have spent more time inside the lossy cavity by the time they are accessed through the reservoir’s noisy readout. Thus, on the longer timescales ($t_S$) at which information is now processed, it is harder for the reservoir (operating at timescale $t_R$) to retain past information. These aspects explain why the overall total memory capacity (summed over all degrees) decreases with increased sample duration $t_S$.
The numerical results on both the linear reservoir ($\gamma=0$) in Figs. \ref{fig:experimentalverification}(e) and (h) and the nonlinear reservoir ($\gamma=\gamma_{Kerr}$)  in Figs. \ref{fig:experimentalverification}(f) and (i) correctly predict a drop in the total linear memory capacities (degrees 1 and 2). Due to the memory capacity cutoff explained in Section 2.5, small capacities are harder to quantify accurately and systematic underestimation can occur. This explains why the small total memory capacities obtained experimentally are larger than the small total memory capacity obtained numerically. The correspondence for large total memory capacities is better as they are largely unaffected by the cutoff. But besides the drop in linear memory capacities, only the nonlinear reservoir model can explain the steady increase in nonlinear memory capacities (degrees 3 and 4) with longer sample durations. With increasing sample duration $t_S$ the simulated nonlinear reservoir shows the contribution of the total nonlinear memory capacity (degrees 3 and 4) to the total memory capacity (all degrees) growing from $0\%$ to $25.4\%$, and in the experiment this contribution starts at $6.4\%$ and grows up to $23.6\%$. This sizable increase in nonlinear computation capacity can be of considerable significance to the reservoir's performance on other tasks, as shown earlier. When comparing the experimental results with the nonlinear reservoir model for all given sample durations $t_S$, the main difference is that the capacities of degree 3 seem to appear sooner (i.e. for smaller sample duration) in the experiment. This can be explained by the residual bias component to the optical injected field. Such a bias makes it easier to produce polynomial functions of odd degrees, thus explaining their earlier onset. This can be explained by the quadratic nature of the Kerr nonlinearity, as the reasoning previously applied to the quadratic nonlinearity of the PD in Eq. \eqref{eq:PD} can be generalized to memory capacities of higher degree. 

\section{Discussion}

We have identified and investigated the role of nonlinear transformation of information inside a photonic computing system based on a passive coherent fiber-ring reservoir. Nonlinearities can occur at different places inside a reservoir computer: the input layer, the bulk and the readout layer. 
State-of-the-art opto-electronic RC systems often include one or several components which inevitably introduce nonlinearities to the computing system. On the reservoir's input side, we have compared a linear input regime with the usage of a MZM, which has a nonlinear transfer function, to convert electronic data to an optical signal. On the reservoir's output side, we have compared a linear output regime with the usage of a PD which measures optical power levels, that scale quadratically with the optical field strength of the neural responses. We numerically evaluated such systems using a benchmark test and found that nonlinear input and/or output components are needed to obtain good RC performance when the optical reservoir itself (i.e. the core of the RC system) is a strictly linear system. 

Internal to the reservoir, we investigated the effect of the optical Kerr nonlinear effect on RC performance. Our numerical benchmark test showed a large band of optical powers where the presence of this distributed nonlinear effect, caused by the waveguiding material of the reservoir, significantly decreased the RC's error figure. Our numerical and experimental measurements of the linear and nonlinear memory capacity of this RC system showed that the accumulation of nonlinear phase due to the distributed nonlinear Kerr effect strongly improves the system's nonlinear computational capacity. We can thus conclude that for photonic reservoir computers with nonlinear input and/or output components, the presence of a distributed nonlinear effect inside the optical reservoir improves the RC performance. Furthermore the distributed nonlinearity is essential for good performance in the regime where nonlineariies are absent from both the input and ouput layer. This may be the case in an all-optical reservoir computer (i.e. with optical input and output layers). We have shown that the effect of the distributed nonlinearity is strong enough to compensate for the lack of nonlinear transformation of information elsewhere in the system, and that it allows to build a computationally strong photonic computing system.

Finally, we expect a design approach including distributed nonlinear effects to improve the scalability of these types of computational devices. In general, when harder tasks are considered, larger reservoirs are required. One way to increase the size of a delay-based reservoir is to implement a longer delay-line. This increase in length of the signal propagation path naturally increases the effect of distributed nonlinearities as considered in this work. Similarly, increasing the size of a network-based reservoir will also lead to more and/or longer signal paths, resulting in the increased accumulation of nonlinear effects, although waveguides with stronger nonlinear effects may have to be considered to compensate for the shorter connection lengths in on-chip implementations. We believe that the natural increase in the strength of nonlinear effects, following the increase in size of the reservoir, may diminish the need to place discrete nonlinear components inside large networks used for strongly nonlinear tasks. As such, both the complexity and cost of such systems would be reduced. Since the waveguiding material itself is used to induce nonlinear effects, the waveguide properties (such as material and geometry) determines the optical field confinement and thus regulate the strength of nonlinear interactions. Consequently it may be possible to create reservoirs where deliberate variations in the waveguide properties are used to tune the strength of the distributed nonlinear effect in different regions of the system. This would allow for a trade off between the system's linear memory capacity and its nonlinear computational capacity, such that a large number of past input samples can be retained (in some parts of the system) and then nonlineary processed to solve difficult tasks (in other parts of the system). These considerations indicate why distributed nonlinear effects may play a major role in future implementations of powerful photonic reservoir computers.

\section*{Conflict of Interest Statement}

The authors declare that the research was conducted in the absence of any commercial or financial relationships that could be construed as a potential conflict of interest.

\section*{Author Contributions}

The idea was first conceived by GVdS and finalized together with GV and SM. JP was responsible for the physical modelling, the numerical calculations and the experimental verification and wrote most of the manuscript. All coauthors contributed to the discussion of the results and writing of the manuscript.

\section*{Funding}

We acknowledge financial support from the Research Foundation Flanders (FWO) under grants 11C9818N, G028618N and G029519N,  the Fonds de la Recherche Scientifique (FRS-FNRS), the Hercules Foundation and the Research Council of the VUB.

\section*{Data Availability Statement}

The data used in this study for the Sante Fe prediction task \cite{weigend1993} is one of the data sets from the “Time Series Prediction Competition” sponsored by the Santa Fe Institute, initiated by Neil Gershenfeld and Andreas Weigend in the early 90’s, no licenses/restrictions apply. No further datasets were used or generated.

\bibliographystyle{frontiersinHLTH&FPHY} 
\bibliography{mybib}


\end{document}